\documentclass[fleqn,10pt]{wlpeerj}

\usepackage{graphicx}
\usepackage{siunitx}

\usepackage[hidelinks]{hyperref}

\newcommand{\Starfishsp}{\emph{Pisaster ochraceus}}
\newcommand{\Oystersp}{\emph{Crassostrea gigas}}
\newcommand{\Clamsp}{\emph{Saxidomas nuttalli}}
\newcommand{\Musselsp}{\emph{Mytilus edulis}}
\newcommand{\Ssp}{\emph{P.~ochraceus}}
\newcommand{\Osp}{\emph{C.~gigas}}
\newcommand{\Csp}{\emph{S.~nuttalli}}
\newcommand{\Msp}{\emph{M.~edulis}}
\newcommand{\Sgen}{\emph{Pisaster}}
\newcommand{\Ogen}{\emph{Crassostrea}}
\newcommand{\Cgen}{\emph{Saxidomas}}
\newcommand{\Mgen}{\emph{Mytilus}}

\newcommand{\Alcyonium}{\emph{Alcyonium}}
\newcommand{\Mycale}{\emph{Mycale}}
\newcommand{\Odontaster}{\emph{Odontaster}}
\newcommand{\Sterechinus}{\emph{Sterechinus}}
\newcommand{\Homaxinellabalfourensis}{\emph{Homaxinella balfourensis}}
\newcommand{\Suberitescaminatus}{\emph{Suberites caminatus}}
\newcommand{\Alcyoniumantarcticum}{\emph{Alcyonium antarcticum}}
\newcommand{\Mycaleacerata}{\emph{Mycale acerata}}
\newcommand{\Odontastervalidus}{\emph{Odontaster validus}}
\newcommand{\Sterechinusneumayeri}{\emph{Sterechinus neumayeri}}

\newcommand{\Adamussiumcolbecki}{\emph{Adamussium colbecki}}
\DeclareSIUnit\frame{frame}
\DeclareSIUnit\psu{psu}

\usepackage{lineno}

\title{Bio-inspired design of ice-retardant devices based on benthic marine invertebrates: the effect of surface texture}

\author[1]{Homayun Mehrabani}
\author[1,4]{Neil Ray}
\author[2]{Kyle Tse}
\author[3,5]{Dennis Evangelista}
\affil[1]{Department of Bioengineering, University of California, Berkeley, CA 94720-1762, USA}
\affil[2]{Department of Mechanical Engineering, University of California, Berkeley, CA 94720-1740, USA}
\affil[3]{Department of Integrative Biology, University of California, Berkeley, CA 94720-3140, USA}
\affil[4]{Current address: Duke University School of Medicine, Durham, NC 27710, USA}
\affil[5]{Current address: Department of Biology, University of North Carolina at Chapel Hill, NC 27510-3280, USA}


\begin{abstract}
Growth of ice on surfaces poses a challenge for both organisms and for devices that come into contact with liquids below the freezing point. Resistance of some organisms to ice formation and growth, either in subtidal environments (e.g.~Antarctic anchor ice), or in environments with moisture and cold air (e.g.~plants, intertidal) begs examination of how this is accomplished.  Several factors may be important in promoting or mitigating ice formation.  As a start, here we examine the effect of surface texture alone. We tested four candidate surfaces, inspired by hard-shelled marine invertebrates and constructed using a three-dimensional printing process. We screened biological and artificial samples for ice formation and accretion in submerged conditions using previous methods, and developed a new test to examine ice formation from surface droplets as might be encountered in environments with moist, cold air. It appears surface texture plays only a small role in delaying the onset of ice formation: a stripe feature (corresponding to patterning found on valves of blue mussels, \Musselsp, or on the spines of the Antarctic sea urchin \Sterechinusneumayeri) slowed ice formation an average of 25\% compared to a grid feature (corresponding to patterning found on sub-polar butterclams, \Clamsp). The geometric dimensions of the features have only a small ($\sim$6\%) effect on ice formation. Surface texture affects ice formation, but does not explain by itself the large variation in ice formation and species-specific ice resistance observed in other work. This suggests future examination of other factors, such as material elastic properties and surface coatings, and their interaction with surface pattern.
\end{abstract}

\begin{document}

\flushbottom
\maketitle
\thispagestyle{empty}

\modulolinenumbers[5]
\linenumbers

\section*{Introduction}
Ice nucleation remains a serious issue for many industrial applications where spray, contact with, or submersion in liquids in sub-freezing point environments is inevitable. Piping systems, ship hulls, airplane wings, and road surfaces are all situations where the accumulation of ice can compromise safety or strength, reduce friction available for traction, impose additional weight, reduce stability, or alter aerodynamic or hydraulic performance.  Though methods of actively removing ice after its formation (chemicals \citep{Hassan:2002}, rubber mallets, inflatable boots, heating cables, bleed air, geothermal \citep{Makkonen:2001} or microwave heating \citep{Hansman:1982} de-icing systems) have been developed to varying degrees of effectiveness, prevention using passive means is preferable \citep{Hassan:2002}.  

In both plants and animals, previous work has shown that ice formation can adversely affect physiology and function or present severe ecomechanical challenges that are likely to be selective \citep{Gusta:2004, Wisniewski:2002, Denny:2011}.  Biological systems may have evolved features to ameliorate the effects of ice that could be copied in engineered devices.  For example, \citep{Gusta:2004} studied ice nucleation within plant leaves and the effects of anti-freeze proteins in systems with internal flow. While such observations are important, they may not be of help in larger scale devices where it is undesirable to alter chemistry; furthermore, extrinsic sources of ice nucleation may be just as damaging \citep{Wisniewski:2002}.  As another example, among benthic marine invertebrates in McMurdo Sound, Antarctica \citep{Dayton:1969, Denny:2011, Mager:2013}, during subcooled conditions, anchor ice readily forms on some organisms (the sponges \Homaxinellabalfourensis\ and \Suberitescaminatus), while others are resistant (the rubbery soft coral \Alcyoniumantarcticum, the sea urchin \Sterechinusneumayeri, the sponge \Mycaleacerata, and the starfish \Odontastervalidus).  Accretion of ice can interfere with physiological function and the resulting positive buoyancy can cause removal; as a result, different suceptibility to ice formation affects the community makeup and local ecology \citep{Dayton:1969, Dayton:1970, Dayton:1989, Pearse:1991, Gutt:2001, Denny:2011, Mager:2013}. Denny et al.~\citep{Denny:2011} did not test the mechanical determinants of the differences in ice formation, though they pointed out the presence of mucus (in \Mycale) in contrast with the others, as well as the difference between spongy or rubbery organisms and those with hard surfaces like the urchins and sea stars.  \citep{Denny:2011} also did not provide contrast with sub-polar organisms.

There are clearly many variables that can affect the formation and adhesion of ice.  Biological systems may make use of many mechanisms of ice inhibition concurrently and may exploit any or all of the following \citep{Gutt:2001, Denny:2011, Gusta:2004, Wisniewski:2002}: shape and mechanical pattern, multi-scale features, internal and surface chemistry, coatings, variation in material properties, animal behavior.  Unfortunately, not all techniques used in living systems are amenable to large-scale engineering application with current technology.  While it is tempting to model everything to the finest scale possible, it is important also to determine the simplest treatment that might work.  Can we narrow the variables of interest, to learn about both the biology and potential applications?  We hypothesized that surface texture provides protection from ice formation, i.e.~for purposes of manufacturability, an attractive engineering solution is one using mechanical patterning alone.  To address if this is possible, we tested four candidate artificial surface textures (grid, valley, cone, stripes) inspired by hard-shelled benthic invertebrates (\Clamsp, \Oystersp, \Starfishsp, and \Musselsp, respectively), with all other mechanical properties (material, stiffness, density, thermal conductivity and specific heat, surface wetting) held constant.  To examine sensitivity, an range of surface parameters (feature spacing and height), bounding the range seen in organisms, was tested.  Tests included both the ice formation test of \citep{Denny:2011} and a new test to examine ice formation in cold air. Such testing is conveniently also the first step in dissecting the underlying biomechanical causes of inter-species variation described in \citep{Denny:2011}; furthermore, it allows comparison between organisms with similar surfaces but different latitudinal distribution (polar versus sub-polar).

\section*{Methods and materials}

\subsection*{Biological samples and ice formation test}
\begin{figure}[t]
\begin{center}
\includegraphics[width=4.875in]{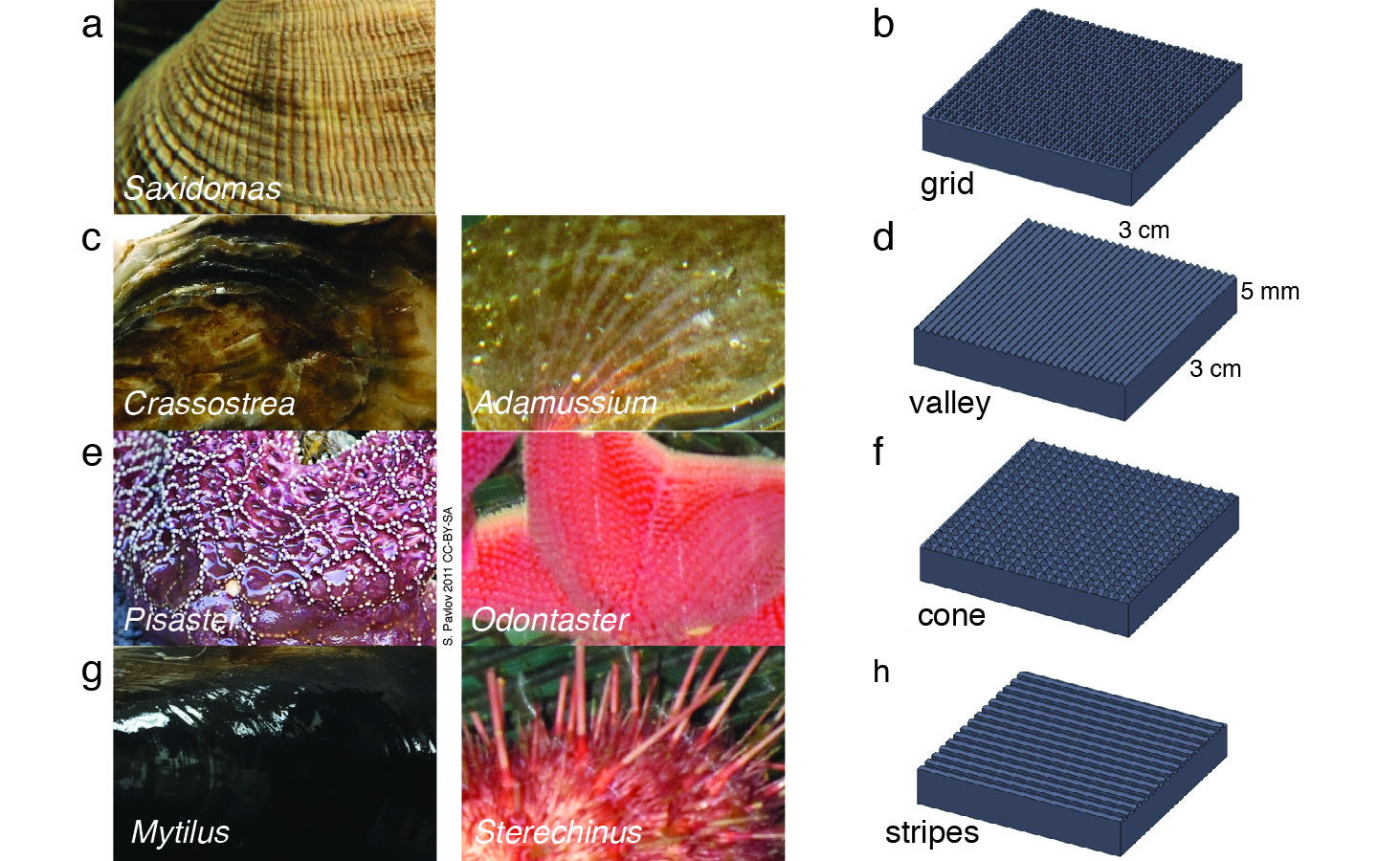}
\end{center}
\caption{Biological inspiration and surface fabrication. (a-b) \Clamsp\ and grid texture; (c-d) \Oystersp\ and valley texture, similar to the Antarctic scallop \Adamussiumcolbecki; (e-f) \Starfishsp\ \citep{Pavlov:2011} and cone texture, similar to the Antarctic starfish \Odontastervalidus; (g-h) \Oystersp\ and striped texture, similar to the patterning on spines of the Antarctic urchin \Sterechinusneumayeri. Photos scaled to approximately \SI{3}{\centi\meter} width.}
\label{fig:bb1}
\end{figure}

We obtained samples of the hard external surfaces of live sub-polar marine invertebrates from a local grocery store (Ranch 99, El Cerrito, CA) and from material leftover from undergraduate biology teaching labs, including butterclams (\Clamsp), blue mussels (\Musselsp), ochre sea star (\Starfishsp), and Pacific oysters (\Oystersp) (Figure~\ref{fig:bb1}). The material was subjected to the ice formation test of \citep{Denny:2011}, to contrast it with the Antarctic species reported by Denny et al.  Biological samples ($<$ \SI{2}{\centi\meter} maximum dimension) were placed in open, unstirred \SI{250}{\milli\liter} polypropylene tripour beakers and filled with \SI{150}{\milli\liter} of \SI{32}{\psu} artificial seawater (Instant Ocean, Blacksburg, VA).  Both sample and seawater were initially at \SI{4}{\celsius}.  Multiple beakers were placed on a fiberglass tray which was then placed in a \SI{-20}{\celsius} walk-in freezer and observed for ice formation on the sample within the first \SI{10}{\minute}.  

\subsection*{Biological inspiration and surface fabrication}
Biological samples were observed under a 10x dissecting microscope, measured and sketched.  To exaggerate specific physical properties in repeatable patterns to observe phenomena that are a result of surface texture alone, we used a three dimensional printing (3DP) process to create engineered samples mimicking the biological samples.  Samples measured \SI{0.03 x 0.03 x 0.005}{\meter} overall and included patterning of four textures: grid, valley, cone, and stripes (Figure~\ref{fig:bb1}). The  textures corresponded to \Cgen, \Ogen, \Sgen, and \Ogen\, respectively.  The cone texture is also similar to the Antarctic starfish \Odontaster, while the striped texture is similar to patterning observed on the spines of the Antarctic urchin \Sterechinus; both are known to be resistant to ice formation \citep{Denny:2011}.  The valley texture is similar to the corrugations on the Antarctic swimming scallop \Adamussiumcolbecki, although such corrugations are likely linked to weight reduction needed in order to swim \citep{Denny:2006}, and the presence of epibionts may alter that organism's propensity to form ice.

The observed textures (Figure~\ref{fig:bb1}) ranged from \SIrange{0.25}{0.5}{\milli\meter} high.  Features were spaced approximately \SIrange{0.5}{0.75}{\milli\meter} for the grid-like texture on \Cgen; approximately \SIrange{1}{3}{\milli\meter} for the groove-like/valley texture on \Ogen; \SIrange{0.5}{4}{\milli\meter} for the cone texture of \Sgen; and \SI{0.5}{\milli\meter} for the stripe texture on \Mgen. These measurements were used in setting designing the textures: for initial screening, a height of \SI{0.5}{\milli\meter} was used, with pattern spacing set at \SI{0.5}{\milli\meter} for grid, valley, and stripes and \SI{1.5}{\milli\meter} for cone. 

We designed the textures using two solid-modeling programs: Solidworks (Dassault Systems, Waltham, MA) and Blender (Blender Foundation, Amsterdam, Netherlands).  Solid models were used to prepare stereolithography (STL) files that were printed in ABS plastic using a ProJet 3000 3D printer (3D Systems, Rose Hill, SC).  Newly printed samples were cleaned of support material by gentle heating and use of an \SI{80}{\celsius} sonicating warm-oil bath, then washed thoroughly with detergent before testing.  

\subsection*{Ice formation test of sample plates}
As with the biological samples, we tested the four sample plates using the ice formation test of \citep{Denny:2011} (Figure~\ref{fig:bb2}(a)).  The onset of ice formation was determined by observing specimens by eye in a walk-in freezer using identical methods to \citep{Denny:2011}.  Based on the results of this test, the best-performing and worst-performing textures were selected for further optimization of the surface parameters (pattern spacing and height), to examine sensitivity. All plates were found to exhibit ice formation during the test period, so further examination of pattern sensitivity used a new test to more closely examine the freeze time as well as ice formation in intertidal or terrestrial cases. 

\subsection*{Droplet tests of sample plates}
\begin{figure}[h]
\begin{center}
\includegraphics[width=3in]{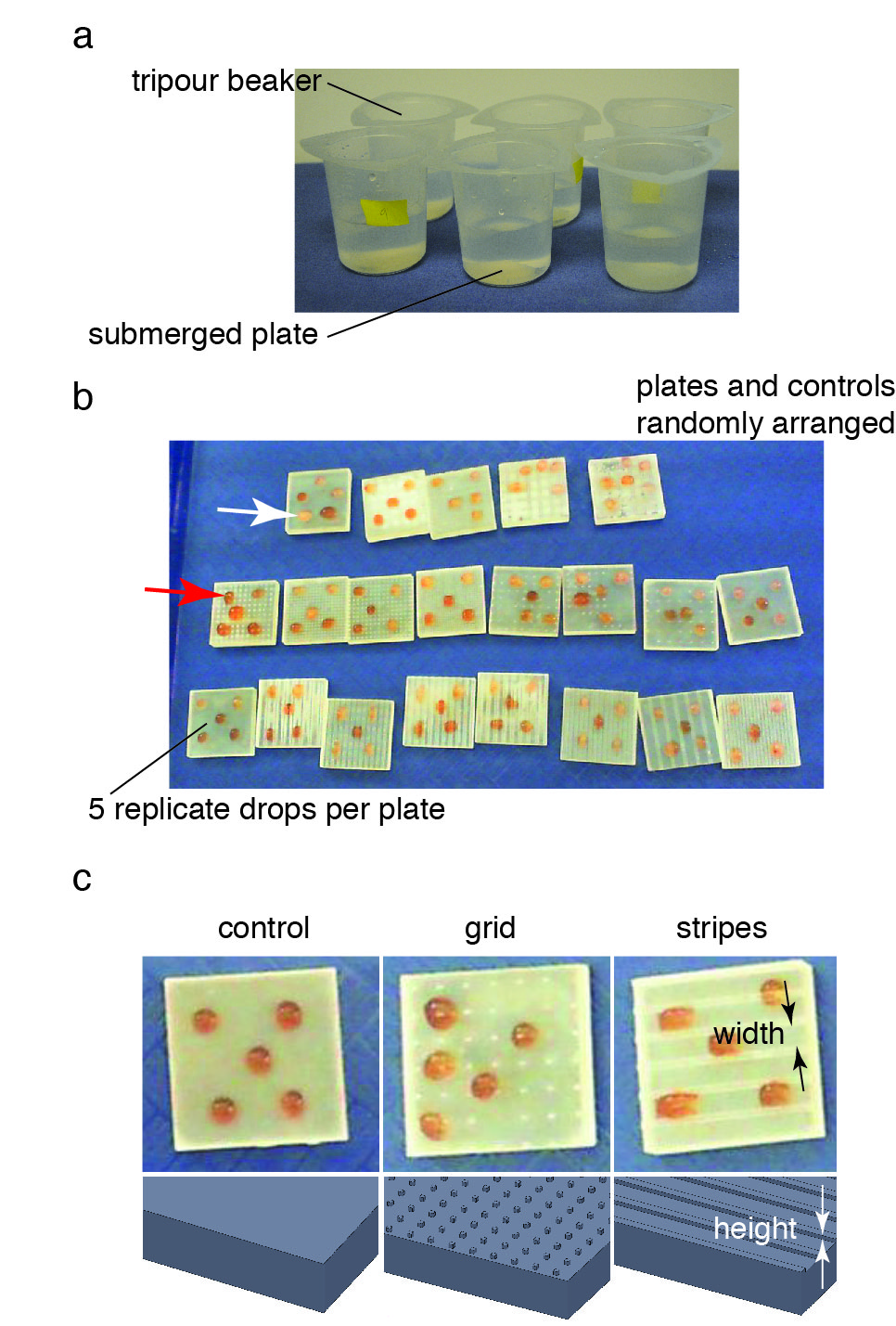}
\end{center}
\caption{(a) Submerged ice formation test from \citep{Denny:2011}.  Plates were placed in \SI{250}{\milli\liter} beakers and watched for ice formation in a \SI{-20}{\celsius} walk-in freezer.  (b) Droplet test to test for ice formation in cold air (intertidal or terrestrial case).  Plates and controls were randomly arranged on a tray in the same \SI{-20}{\celsius} freezer. Droplet freezing was identified by color shift from red (middle arrow) to white (upper arrow).  Plates \SI{0.03}{\meter} square. (c) Textures and control used during droplet test.  Feature width and height varied between \SIrange{0.5}{4}{\milli\meter} and \SIrange{0.25}{1}{\milli\meter} respectively.}
\label{fig:bb2}
\end{figure}

An additional series of sample plates based on the best-performing and worst-performing textures from the ice formation test was prepared, varying the pattern spacing from \SIrange{0.5}{4}{\milli\meter} and pattern height from \SIrange{0.25}{1}{\milli\meter}. In addition, several untextured control plates were also prepared (STL files available for download).  Plates were dried and stored in \SI{4}{\celsius} walk-in freezer overnight to provide isothermal starting conditions. Plates were placed randomly in rows on a plastic tray for testing.  Five \SI{0.1}{\milli\liter} droplets of \SI{32}{\psu} artificial seawater, also at \SI{4}{\celsius}, were placed on the four corners of the plate and the center (Figure~\ref{fig:bb2}(b-c)). The plates were then placed in a \SI{-20}{\celsius} freezer on a flat surface away from overhead fans and observed during freezing.  In addition to visual observation of samples with the naked eye, a digital video camera (Hewlett Packard, Palo Alto, CA) was used to obtain time lapse images at \SI{1}{\frame \per\second} (example video available for download). Samples were observed over \SI{30}{\minute} or until all of the samples froze. Samples were classified as frozen once the droplet turned opaque white (Figure~\ref{fig:bb2}(b)). The resulting freeze times were recorded and analyzed in R \citep{R:2014}.  For the results presented below, freeze times for a given test were normalized by the mean freeze time for the untextured control plates in the test. 

\section*{Results}

\subsection*{Ice formation test for biological samples and sample plates}
\begin{figure}
\begin{center}
\includegraphics{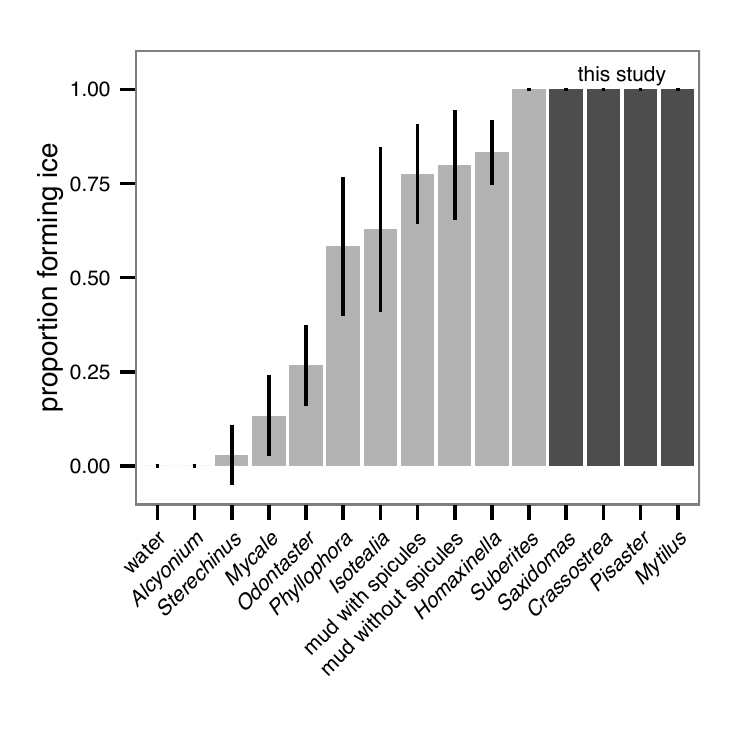}
\end{center}
\caption{Ice formation on sub-polar organisms (\Clamsp, \Oystersp, \Starfishsp, and \Musselsp, dark gray) compared to Antarctic data reproduced from \citep{Denny:2011} (light gray).  All sub-polar samples tested here initiated ice formation during the ice formation test (3 batches, $n=5$ each).}
\label{fig:dennycomparison}
\end{figure}

Unlike in Antarctic species tested in \citep{Denny:2011}, all sub-polar biological samples we tested (hard external surfaces of \Clamsp, \Oystersp, \Starfishsp, and \Musselsp) initiated ice formation during the ice formation test (Figure~\ref{fig:dennycomparison}).  This is discussed further below. 

\begin{figure}
\begin{center}
\includegraphics{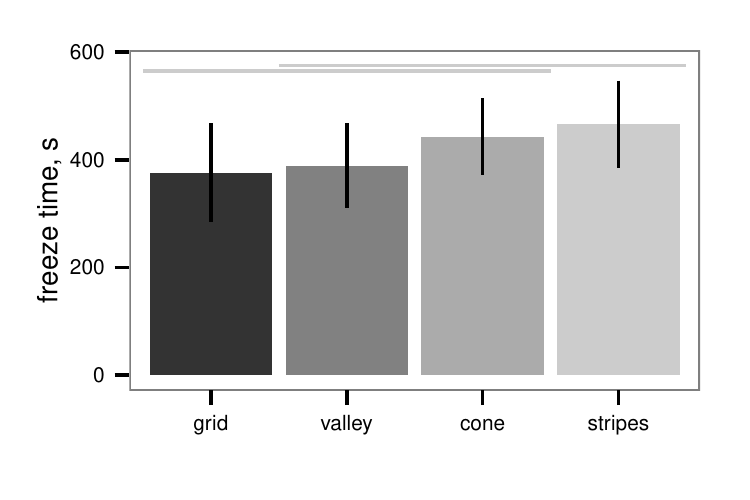}
\end{center}
\caption{Freeze time (mean $\pm$ s.d.) for sample plates during ice formation test. Differences between textures are significant (ANOVA, $P=0.019$); light grey lines indicate groups from post-hoc Tukey analysis.}
\label{fig:bb3}
\end{figure}

All sample plates tested in the submerged ice formation test of \citep{Denny:2011} also initiated ice formation, however, there were broad differences in freeze time between the textures (ANOVA, $P=0.019$, Figure~\ref{fig:bb3}).  The best-performing texture (stripes) increased the time to freeze relative to the worst-performing texture (grid), by 25\%.  On the other hand, wetted surface area did not appear to affect the freeze time (linear regression, $P=0.223$, Table~\ref{tbl:bb3}).  

\begin{table}
\caption{Freeze time (mean $\pm$ s.d.) and wetted area for sample plates during submerged ice formation test. Freeze time does not depend on wetted area (linear regression, $P=0.223$).}
\label{tbl:bb3}
\begin{center}
\begin{tabular}{rcc}
\hline
texture & freeze time, \SI{}{\second} & area, \SI{}{\milli\meter\squared} \\
\hline
grid & $375\pm92$ & 1800 \\
valley & $388\pm79$ & 1290 \\
cone & $433\pm71$ & 1030 \\
stripes & $465\pm81$ & 1520 \\
\hline
\end{tabular}
\end{center}
\end{table}

\subsection*{Droplet test and sensitivity of freeze time to pattern spacing}
\begin{figure}
\begin{center}
\includegraphics{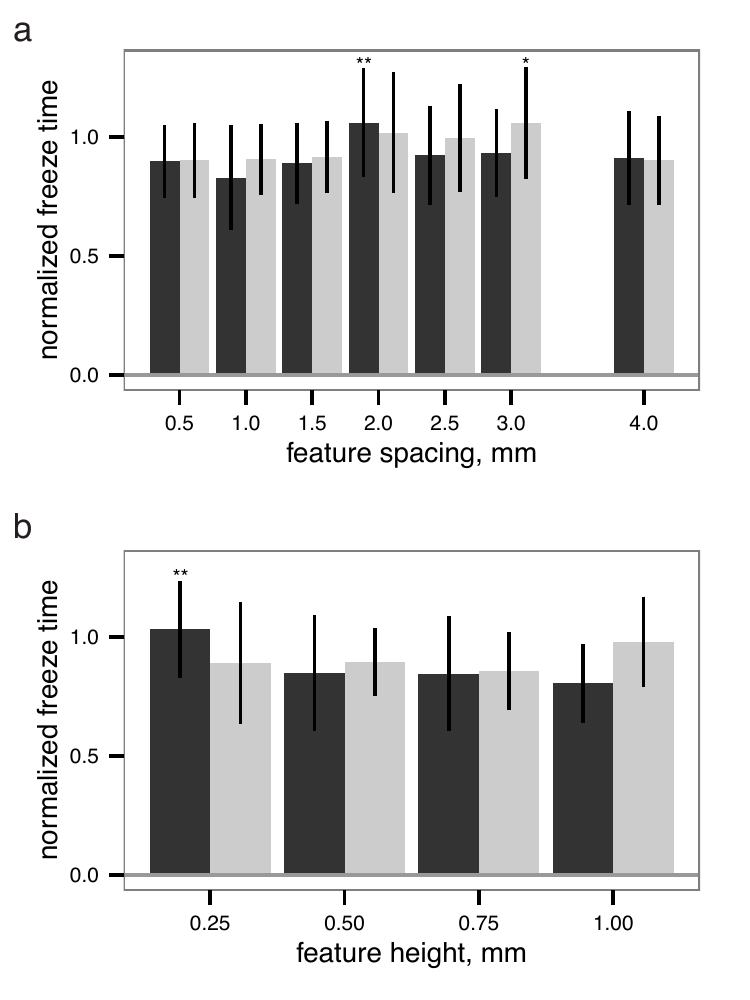}
\end{center}
\caption{Normalized freeze time versus (a) feature spacing and (b) feature height (see Figure~\ref{fig:bb2}(c)), examining sensitivity to surface parameters for two different textures. Grid texture in dark grey; stripes texture in light grey as in Figure~\ref{fig:bb3}.  Maxima indicated by * for stripes and ** for grid; normalized freeze time does depend on surface parameters (ANOVA, $P=0.0004$ for spacing, $P=0.0389$ for height), but increases in freeze time are small relative to the noise in the measurement.}
\label{fig:bb4}
\end{figure}

Within both the best-performing texture (stripes) and the worst-performing texture (grid), the pattern spacing appeared to alter the normalized freeze time for droplets up to 28\%, though the slowest freeze times were only delayed about 6\% relative to the untextured control (Figure~\ref{fig:bb4}).  Given a texture, there appears to be a (weakly) optimal feature spacing (ANOVA, $P=0.0004$ for spacing, $P=0.0389$ for height), however, increases in normalized freeze time obtained by varying pattern spacing or height are small relative to the noise in the measurement.  An example time-lapse movie of a droplet test is provided in the supplemental material.   

\section*{Discussion}

\subsection*{Sub-polar species always initiated ice formation}
It is perhaps not a surprise that the four species tested (\Clamsp, \Oystersp, \Starfishsp, and \Musselsp) all initiated ice formation (Figure~\ref{fig:dennycomparison}). Although all are found in cold water (\Csp, \Ssp, and \Osp\ range to Alaska), they are not particularly noteworthy as polar species.  On the other hand, hard shelled bivalves of closely-related species can be major components of Arctic communities \citep{Dayton:1994, Gutt:2001}, while echinoderms (e.g.~\Odontaster\ and \Sterechinus) are hugely important in the Antarctic \citep{Dayton:1969, Dayton:1994}.  The relative ease with which the sub-polar species here initiated ice formation, compared to Antarctic species reported in \citep{Denny:2011}, suggests one potential barrier to polar spread: even mild anchor ice events would likely remove all of the sub-polar species we tested.  While \Msp\ is tolerant of being frozen in ice \citep{Aarset:1982, Aunaas:1985, Kanwisher:1955}, \citep{Gutt:2001} notes that glacier scour and buoyancy effects of ice can cause removal.  Unfortunately, as ice recedes, such barriers may be diminishing.

\subsection*{Implications for device design}
Contrary to what we hypothesized, simple surface texture does not appear to be a magic bullet able to confer a large degree of ice resistance (Figures~\ref{fig:bb3} and \ref{fig:bb4}).  However, careful application of a correctly-spaced texture appears to delay the onset of freezing a small amount, though the mechanism is unclear.  For droplet tests in air, one mechanism that may explain the observed maxima is the interaction between hydrophobicity conferred by finely spaced textures \citep{Cassie:1948} and insulation from entrained air within the texture. While our tests did not include measurement of contact angles, the ratio of pattern spacing used in our tests corresponded to the region where surface patterning can increase apparent contact angle \citep{Cassie:1948}, potentially increasing the thermal insulation provided by trapped air on one side of the drop compared to thermal conduction through the features.  However, such a mechanism and its interaction with thermal conduction is uncertain since the minima in these noisy data fall at intermediate spacing and low height.  Biologically, while a Cassie wetting mechanism cannot explain subtidal differences as observed in \citep{Denny:2011}, it could be relevant in the feather or fur coats of animals, in the intertidal, or in trichomes on leaves, perhaps causing extracorporeal ice nucleation as a means to provide insulation to more sensitive tissues beneath \citep{Duman:1991}. 

Surface chemistry may have effects on ice resistance; hydrophilic materials reduce resistance to ice adhesion \citep{Meuler:2010, Meuler:2010b} in air, and many biological materials (notably calcium carbonate in bivalve shells and echinoderm ossicles and spines) are hydrophilic.  It is unclear what the effect of surface chemistry is in subtidal situations, and the ice resistance of \Sterechinus\ and \Odontaster\ (both echinoderms with calcareous ossicles) appears contrary to expectations from surface chemistry alone.

\subsection*{Texture has demonstrable, but small, effects on freezing time}
While texture does have demonstrable effects on freezing time, the effects are small (Figures~\ref{fig:bb3} and \ref{fig:bb4}).  Mechanical patterning alone cannot explain the large inter-specific differences in ice formation observed in \citep{Denny:2011} for Antarctic species, or the ease with which sub-polar species appear to initiate ice formation. Thus, we infer that the differences observed by \citep{Denny:2011} are due to other effects or interactions with surface texture.  

Further dimensions that may be exploited in biological systems but were not examined here include surface wetting (hydrophobic/hydrophilic); surfactants, mucus and slime (as in \Mycale); and elastic and viscoelastic properties (e.g.~rubbery \Alcyonium); nanoscale pattern, multi-scale shape and behavior. In an engineering system, these would increase cost; however, perhaps a subset of these may provide synergistic effects resulting in performance improvement in excess of the cost increase.  For example, the combination of rigid and viscoelastic materials in a micro-scale composite artificial shark skin was effective in drag reduction \citep{Wen:2014}.  For ice retardant devices, clearly mechanical designs patterned solely on sub-polar species have not yet provided performance of the same level as the actual polar organisms.  Additional observation is needed of multi-scale shape, behavior, and the material properties of perishable tissue and secretions that do not preserve well.  Further investigation of ice biology in the field, to examine biomechanical performance in natural environments, along with innovative pairing of materials may enable more effective anti-ice engineering solutions.

\section*{Acknowledgements}
We thank T Libby and the Berkeley Center for Integrative Biomechanics Education and Research (CIBER); A Doban for use of a \SI{-20}{\celsius} freezer for testing of devices; E Kepkep for providing invertebrate samples; and R Dudley and T Hedrick for their support.  We thank two anonymous reviewers for comments which improved the manuscript.  This work was supported by the UC Berkeley Undergraduate Research Apprenticeship Program (URAP). 

\bibliography{frostyboy}
\end{document}